\documentstyle[12pt,titlepage]{article}

\font\grande=cmr9.5 scaled \magstep4
\font\medio=cmr9.5 scaled \magstep2
\outer\def\beginsection#1\par{\medbreak\bigskip
      \message{#1}\leftline{\bf#1}\nobreak\medskip
\vskip-\parskip
      \noindent}

\def\laq{\raise 0.4ex\hbox{$<$}\kern -0.8em\lower 0.62
ex\hbox{$\sim$}}
\def\gaq{\raise 0.4ex\hbox{$>$}\kern -0.7em\lower 0.62
ex\hbox{$\sim$}}

\begin{document}

\titlepage
\begin{flushright}
CERN-PH-TH/2005-085
\end{flushright}

\vspace{15mm}

\begin{center}

{\grande Inhomogeneous dusty Universes and their deceleration}

\vspace{15mm}
\large{Massimo Giovannini \footnote{e-mail address: massimo.giovannini@cern.ch}}
\end{center}
\vspace{6mm}
\centerline{\medio{{\sl Centro ``Enrico Fermi", Compendio del Viminale, Via 
Panisperna 89/A, 00184 Rome, Italy}}}
\vspace{6mm}
\centerline{\medio{{\sl Department of Physics, Theory Division, CERN, 1211 Geneva 23, Switzerland}}}
\vskip 2cm

\centerline{\medio Abstract}

\noindent
Exact results stemming directly from Einstein equations 
imply that inhomogeneous Universes endowed with vanishing 
pressure density can only decelerate, unless the energy density of the Universe becomes negative.
 Recent proposals seem to argue that inhomogeneous (but isotropic) space-times, filled only with incoherent matter,
may turn into accelerated Universes for sufficiently late times.
To scrutinize these scenarios, fully inhomogeneous Einstein equations are 
discussed in the synchronous system. In a dust-dominated Universe, the inhomogeneous generalization of the 
deceleration parameter is always positive semi-definite implying that no acceleration takes place.
\vspace{5mm}
\vfill
\newpage
The homogeneity and isotropy of the geometry are not essential ingredients 
to establish a number of relevant results in relativistic cosmology.
For instance, from the early sixties to the early seventies (see \cite{LK1,BLK,KL} and references therein),
a research program on the singularity properties of general 
cosmological solutions has been conducted without relying on the 
isotropy and on the homogeneity of the geometry.

The theme of the 
present paper is somehow opposite to the one 
analysed in \cite{LK1,BLK,KL} where the emphasis 
was on the r\^ole of the inhomogeneities (and anisotropies) 
in the proximity of a cosmological singularity. In the present paper 
we would like to understand if an inhomogeneous space-time, filled with 
incoherent matter, can be turned into an accelerating Universe at later times.  
The inhomogeneities considered in the present investigation
 may arise during an early inflationary stage when quantum mechanical fluctuations of the geometry and of the inflaton field
are inside the Hubble radius. Depending upon the parameters of the inflationary phase, the initial quantum fluctuations 
will be amplified leading to a quasi-flat spectrum of curvature 
perturbations that accounts, through the Sachs-Wolfe effect, for  the tiny 
temperature ripples detected in the microwave sky by several experiments. While the temperature 
of the Cosmic Microwave Background (CMB) is in the range 
of few ${\rm K}$ degrees, the inhomogeneities are in the 
range of the $\mu {\rm K}$. This hierarchy of scales also implies 
that curvature fluctuations, say, after equality and before decoupling, 
are rather well described in the framework of linear theory.

In recent months, it has been claimed \cite{SH} 
that since spatial inhomogeneities may have a non-trivial time dependence on 
length-scales larger than the Hubble radius, then, depending on the 
specific properties of the inhomogeneity, the deceleration parameter 
may well be negative, implying an effective acceleration of the Universe.
While this proposal has been already criticized on various grounds
\cite{GCA,Fla,selhir,RAS,siegel} we find appropriate to scrutinize 
such a statement in the light of a set of fully inhomogeneous Einstein 
equations. Consider, indeed, Einstein equations 
\begin{equation}
R_{\mu}^{\nu} = 8\pi G \,\biggl[ T_{\mu}^{\nu} - \frac{1}{2} \delta_{\mu}^{\nu} \,T \biggr],\,\,\,\,\,\,\,\,\,\,\,\, T = T_{\mu}^{\mu},
\label{E1}
\end{equation}
where $R_{\mu}^{\nu}$ is the Ricci tensor. 
In the specific case of a perfect relativistic fluid
\begin{equation}
T_{\mu}^{\nu} = ( p + \rho) u_{\mu} u^{\nu} - p \delta_{\mu}^{\nu},
\,\,\,\,\,\,\,\,\,\,\,\, T_{\mu}^{\mu}= (\rho - 3 p),
\label{T1}
\end{equation}
where $u_{\mu}$ is the velocity field obeying
\begin{equation}
g^{\mu\nu} u_{\mu} u_{\nu} =1,
\label{T2}
\end{equation}
while $g^{\mu\nu}$ is the inverse of the four-dimensional metric 
tensor.  The form of $T_{\mu}^{\nu}$ given in Eq. (\ref{T1})  (together with the synchronous condition) excludes the presence of torque forces. 

Being $g_{\mu\nu}$ a symmetric rank-two tensor in $4$ dimensions, 
it contains $10$ independent entries, but $4$ of them 
are associated  with the freedom of choosing a coordinate system.
Without loss of generality,  the synchronous form of the line element may be adopted
\begin{equation}
ds^2 = g_{\mu\nu}(\vec{x}, t) dx^{\mu} dx^{\nu} = 
 dt^2 - \gamma_{ij}(\vec{x},t) dx^{i} \,dx^{j},
\label{metric}
\end{equation}
where the symmetric 3-dimensional tensor $\gamma_{ij}(\vec{x},t)$  appearing in Eq. (\ref{metric}) carries $6$ degrees of freedom which also correspond to the correct number of initial conditions for the general treatment of the  problem. Equation (\ref{metric})  allows the determination of the 
components of the extrinsic and intrinsic (i.e. spatial) curvatures, 
\begin{eqnarray}
&& K_{i}^{j} = - \frac{1}{2} \gamma^{i k} \frac{\partial }{\partial t} \gamma_{k j},
\,\,\, \,\,\,\,\,\,\,\,\,\,\,\,
K= K_{i}^{i},\,\,\,\,\,\,\,\,\,\,\,\,{\rm Tr}K^2 = K_{i}^{j} K_{j}^{i},
\label{Kij}\\
&& r_{i j} = \partial_{m} \Gamma^{m}_{i j} - \partial_{j} \Gamma^{m}_{m i} + 
\Gamma_{i j}^{m} \Gamma_{m \ell}^{\ell} - \Gamma_{j m}^{\ell} \Gamma_{i \ell}^{m},
\label{rim}
\end{eqnarray}
where the three-dimensional Christoffel symbols are constructed 
using directly $\gamma_{ij}$.
The $(00)$, $(ij)$ and $(i0)$ components of Eq. (\ref{E1}) become, respectively:
\begin{eqnarray}
&& \dot{K} - {\rm Tr} K^2 = 8\pi G\biggl[ (p+ \rho) u_{0}\,u^{0} +
\frac{p -\rho}{2} \biggr], 
\label{E200}\\
&& \frac{1}{\sqrt{\gamma}} \frac{\partial}{\partial t}\biggl(\sqrt{\gamma} \,K_{i}^{j}\biggr) - r_{i}^{j} = 8\pi G \biggl[ - (p+ \rho) u_{i} u^{j} 
+ \frac{p -\rho}{2} \delta_{i}^{j} \biggr],
\label{E2ij}\\
&& \nabla_{i} K - \nabla_{k} K^{k}_{i} = 8\pi G (p + \rho) u_{i}\,u^{0},
\label{E20i}
\end{eqnarray}
where $r_{i}^{j} = \gamma^{k j}\,r_{k i}$; the overdot denotes a 
 partial derivation with respect to $t$ while $\nabla_{i}$ denotes 
 the covariant derivative defined in terms of $\gamma_{ij}$; in Eq. (\ref{E2ij}) 
 $\gamma = {\rm det} \gamma_{i j}$ satisfying $ 2 K = - \dot{\gamma}/\gamma$.
Equations (\ref{E200}), (\ref{E2ij}) and (\ref{E20i}) have been used, in a 
slightly different form, in \cite{LK1,BLK,KL}
to scrutinize the properties of inhomogeneous (and anisotropic) Universes. 
 
Using Eq. (\ref{metric}),   Eq. (\ref{T2}) becomes
\begin{equation}
u_{0}u^{0} = 1 + \gamma^{i j} u_{i} u_{j} \geq 1,
\label{cond0}
\end{equation}
where the second inequality holds since $\gamma^{ij}\geq 0$.
By definition of $K$ and ${\rm Tr} K^2$ it is clear that 
\begin{equation}
{\rm Tr}K^2 \geq \frac{1}{3} K^2 \geq 0,
\label{cond1}
\end{equation}
where the first equality is reached in the isotropic case.
Finally, the general definition of the deceleration parameter 
can be easily read-off from Eq. (\ref{E200}) and it is 
\begin{equation}
q(\vec{x},t) = \frac{  \dot{K} - {\rm Tr} K^2 }{{\rm Tr} K^2} \equiv - 1 + \frac{\dot{K}}{{\rm Tr} K^2}.
\label{defq}
\end{equation}
In the isotropic limit $\gamma(\vec{x}, t) \to a^2(t) \delta_{ij}$, 
$K_{i}^{j} \to - (\dot{a}/a) \delta_{i}^{j}$  and, from Eq. (\ref{defq})  $q(\vec{x},t)\to - \ddot{a} a/\dot{a}^2$. Given the results 
reported in Eqs. (\ref{cond0}) and (\ref{cond1}), 
Eq. (\ref{E200}) implies that, since $u_{0} u^{0} \geq 1$, 
  \begin{equation}
  \dot{K} - {\rm Tr} K^2 = 8\pi G\biggl[ (p+ \rho) u_{0}\,u^{0} +
\frac{p -\rho}{2} \biggr]  \geq 4\pi G ( 3p + \rho),
\label{COND3}
\end{equation}
implying, together with Eq. (\ref{cond1}), that in the 
fully inhomogeneous case $ q(\vec{x},t) \geq 0$  provided $(3 p + \rho) > 0$. If $(3p +\rho)<0$ the matter content of the Universe violates the strong energy condition.  No assumption has been made on the geometry 
as being split into a dominant background supplemented by a (comparatively 
small) inhomogeneous perturbation. 

Assuming now (as done in \cite{SH}) that the matter content of the Universe is provided by a dusty fluid with $p=0$, the Universe can accelerate (i.e. $q(\vec{x},t)<0$)  only if
the energy density $\rho$  becomes negative ( violating simultaneously the 
weak, strong and dominant energy consitions).  By taking the difference between Eq. (\ref{E200}) and the trace of Eq. (\ref{E2ij}) the standard form 
of the Hamiltonian constraint can be obtained, in the case $p=0$,
\begin{equation}
K^2 - {\rm Tr}K^2 + r = 16 \pi G \rho \,u_{0} \,u^{0},
\label{HAM}
\end{equation}
i.e. the negativity of the energy density  also implies, according to Eq. (\ref{HAM}),  that $K^2 - {\rm Tr}K^2 + r <0$.
 
Before proceeding further, it is appropriate to note that Eqs. (\ref{E200}), (\ref{E2ij}) and 
(\ref{E20i}) are sufficient to fully determine the dynamical evolution. 
However,  it is always wise to keep an eye on the 
exact form of the covariant conservation equation 
whose two components become, in the case $p =0$
\begin{eqnarray}
&& \frac{1}{\sqrt{\gamma}}
\frac{\partial}{\partial t} (\sqrt{\gamma} \,\rho \,u_{0}\, u^{0} ) + 
\frac{1}{\sqrt{\gamma}} \partial_{i} (\sqrt{\gamma} \,\rho\, u_{0} \,u^{i} )
-  \rho\, K_{i k} u^{k} u^{i}  =0,
\label{COV0}\\
&& \frac{1}{\sqrt{\gamma}} \frac{\partial}{\partial t}[ \sqrt{\gamma}  \rho u^{0} u^{i}] + \frac{1}{\sqrt{\gamma}}\partial_{k} ( \sqrt{\gamma} \rho u^{k} u^{i} ) 
- 2\,\rho\, K_{k}^{i}   u^{0} \,u^{k} + 
\rho\,\Gamma^{i}_{k \ell}   u^{k} \,u^{\ell}  =0.
\label{COVi}
\end{eqnarray}
Every arbitrarily complicated solution of Eqs. (\ref{E200}), 
(\ref{E2ij}) and (\ref{E20i}) must also satisfy, for consistency, 
Eqs. (\ref{COV0}) and (\ref{COVi}).

We are now in condition of discussing some possible solutions 
of our system with the aim of showing that spatial inhomogeneities 
may turn the deceleration parameter from positive to negative. To 
do so the spatial inhomogeneities should be rather strong, so it will 
not be appropriate to separate the geometry into a homogeneous 
component supplemented by some small fluctuation. Hence, 
following  Ref. \cite{SH} we can assume 
\begin{equation}
\gamma_{ij} (\vec{x},t) = a^2(t) e^{ - 2 \Psi(\vec{x},t)} \delta_{ij},
\label{par1}
\end{equation}
that implies, according to Eq. (\ref{Kij}), 
\begin{equation}
K_{i}^{j} = - (H - \dot{\Psi}) \, \delta_{i}^{j}, \,\,\,\,\,\,\,
K = - 3 (H-\dot{\Psi}),\,\,\,\,\,\, {\rm Tr} K^2 = 3 (H - \dot{\Psi})^2.
\label{ext}
\end{equation}
From Eq. (\ref{rim}), $r_{i}^{j}$ and $r$ are instead:
\begin{eqnarray}
&& r_{i}^{j} = \frac{e^{2\Psi}}{a^2} [ \nabla^2 \Psi + 
\partial_{i}\partial^{j} \Psi - (\nabla \Psi)^2 \delta_{i}^{j} + \partial_{i}\Psi
\partial^{j} \Psi ],
\label{in1}\\
&& r= \frac{e^{2\Psi}}{a^2} [ 4 \nabla^2\Psi - 2 (\nabla\Psi)^2],
\label{in2} 
\end{eqnarray}
where $\nabla^2 \Psi= \delta^{ij}\,\partial_{i}\partial_{j}\Psi$ and 
$(\nabla\Psi)^2= \delta^{ij} \,\partial_{i} \Psi \partial_{j} \Psi$.
Equations (\ref{E200}), (\ref{E20i}) and (\ref{HAM}) lead then to the following 
system:
\begin{eqnarray}
&& 
3 \ddot{\Psi} - 3 \dot{\Psi}^2 + 6 H \dot{\Psi} = 4\pi G \biggl[ \tilde{\rho}+ 2 (\overline{\rho} +\tilde{\rho}) \frac{e^{2\Psi}}{a^2} u^2\biggr],
\label{00b}\\
&& \partial_{i} \dot{\Psi} = 4\pi G (\overline{\rho} + \tilde{\rho}) u_{i}
\sqrt{1 + \frac{e^{2\Psi}}{a^2}\,u^2},
\label{momb}\\
&& \dot{\Psi}^2 - 2 H\dot{\Psi} + 
\frac{e^{2 \Psi}}{3 a^2} [ 2 \nabla^2 \Psi - (\nabla\Psi)^2] = 
\frac{8\pi G}{3}\biggl[ \tilde{\rho} + (\overline{\rho} + \tilde{\rho}) \frac{e^{ 2 \Psi}}{a^2} u^2 \biggr],
\label{HAMb}
\end{eqnarray}
where $u^2 =\delta^{ij}\, u_{i} \,u_{j} $ and where the energy 
density has been separated, for later convenience, as 
$\rho = \overline{\rho} + \tilde{\rho}$ with $\overline{\rho}$ satisfying 
the usual Friedmann equations, i.e. 
\begin{equation}
H^2 = \frac{8\pi G}{3} \overline{\rho},\,\,\,\,\,\, \dot{H} +H^2 = - \frac{4\pi G}{3} \overline{\rho},
\label{FRW}
\end{equation}
and implying $2\dot{H} = - 3 H^2$. Eliminating one of the $\tilde{\rho}$ from Eq. (\ref{00b}) through Eq. (\ref{HAM}) the equation to be solved becomes
\begin{equation}
2 \ddot{\Psi} - 3 \dot{\Psi}^2 + 6 H\dot{\Psi} - \frac{e^{2 \Psi}}{3 \,a^2} [ 2 \nabla^2 \Psi - (\nabla\Psi)^2] = \frac{8\pi G}{3} (\overline{\rho} + \tilde{\rho}) 
\frac{e^{2 \Psi}}{a^2} u^2. 
\label{dec}
\end{equation}
We can now look for a solution of the system in the form 
\begin{equation}
\Psi(\vec{x},t) = f+ \lambda(t) e^{2 f} \nabla^2 f, 
\label{ans}
\end{equation}
where $f=f(\vec{x})$ encodes 
the information on the large-scale inhomogeneities of inflationary origin and where $e^{2 f} \nabla^2 f$ is the expansion parameter . In this 
expansion terms like $e^{2 f}(\nabla f)^2$ will be considered subleading 
in comparison with $e^{2 f}\nabla^2 f$, i.e. $|\nabla^2 f| \gg (\nabla f)^2$.
Within this ansatz the momentum constraint can be solved, to lowest 
order as 
\begin{equation}
u_{i} = \frac{\partial_{i} \dot{\Psi}}{4\pi G( \overline{\rho} + \tilde{\rho})}.
\end{equation}
Plugging this solution back into Eq. (\ref{dec}) we find that the term 
containing $u^2$ is of higher order. 
The term $\dot{\Psi}^2$ is also of higher 
order since it contains $\dot{\lambda}^2 e^{4 f} (\nabla^2 f)^2$.
The remaining terms cancel provided 
\begin{equation}
\ddot{\lambda} + 3 H \dot{\lambda} = \frac{1}{3 a^2}.
\end{equation}
If we now normalize the scale factor as $a(t) = (t/t_{0})^{2/3}$ (where $t_{0}$ is the present time) $\lambda(t)$ is easily determined to be 
\begin{equation}
\lambda(t) = \frac{3}{10} t_{0}^2 a(t)= \frac{2}{15\, H_{0}^2} a(t),
\label{sol1a}
\end{equation}
where, according to Eq. (\ref{FRW}), we used $H_{0} = 2/(3\,t_{0})$. 
 The solution given in Eqs. (\ref{ans}) and (\ref{sol1a}) 
coincides with the one 
given in Ref. \cite{SH}. Since 
terms $(\nabla f)^2$ have been neglected in comparison with $\nabla^2 f$,
 a fortiori, in Eq. (\ref{dec}) the term $3\dot{\Psi}^2$ is 
subleading with respect to $2 \ddot{\Psi}$. In fact $\dot{\Psi}^2$ is 
of order $\dot{\lambda}^2 e^{4 f} (\nabla^2 f)^2$ while $\ddot{\Psi}$ is 
of order $\ddot{\lambda} e^{ 2 f} \nabla^2 f$.

The obtained solution can now be parametrized as 
\begin{equation}
\Psi = f(\vec{x}) + a(t) \overline{\Psi}_{0}(\vec{x}).
\label{par}
\end{equation}  
where the time-independent quantity $\overline{\Psi}_{0}$ is effectively treated, 
by the authors of Ref. \cite{SH}, as a free parameter.  Inserting Eq. (\ref{par}) 
into Eq. (\ref{defq}) we obtain 
\begin{equation}
q(\vec{x},t) = -1 + \frac{3/2 - a(t) \overline{\Psi}_{0}/2}{( 1 - a(t) \overline{\Psi}_{0})^2},
\label{decSH}
\end{equation}
that coincides with the expression of the deceleration parameter given in 
\cite{SH}.

From Eq. (\ref{decSH}), the authors of Ref. \cite{SH} conclude  that :
{\it (i)} if $\overline{\Psi}_{0}=0$, then $q(\vec{x},t)= 1/2$;
{\it (ii)} if $|\overline{\Psi}_{0}|< 1$ (and $\overline{\Psi}_{0} <0$)
$q(\vec{x},t) <0$; {\it (iii)} for large $a(t)$ (i.e. $t\to +\infty$) $q(\vec{x}, t)\to -1$.
The only correct statement is {(\it i)}.  The other two are
incompatible with the approximations made in solving both the
dynamical equations and the constraints.

In simple terms, the reason is the following.
The  deceleration parameter given in Eq. (\ref{decSH}) 
 is derived from Eq. (\ref{defq}) using the fact that 
${\rm Tr}K^2 = 3 (H - \dot{\Psi})^2 = 3 H^2 ( 1 - a \overline{\Psi}_{0})^2$:
but this is not correct since terms $\dot{\Psi}^2$ have been 
neglected in the solution (because of higher order).  
Thus,  if we treat $\overline{\Psi}_{0}$ as a free parameter 
we have to make sure, in particular, that from Eq. (\ref{dec}) the term
$2 \ddot{\Psi}$  is {\em always } leading with respect to $3\dot{\Psi}^2$:
using the parametrization (\ref{par}) and the relation $2 \dot{H} = - 3 H^2$ the requirement $ 2 |\ddot{\Psi}| \geq 3 |\dot{\Psi}|^2 $ implies 
\begin{equation}
|a(t) \overline{\Psi}_{0}| \leq \frac{1}{3},
\label{limit}
\end{equation} 
that is the most restrictive condition.
If we now take into account that $\overline{\Psi}_{0}= 
|e^{2 f} \nabla^2 f |<1 $ and that $\overline{\Psi}_{0} < 0$, 
Eq. (\ref{limit}) implies an upper limit on $a(t)$, i.e. 
\begin{equation}
a(t) \leq \frac{1}{3 |\overline{\Psi}_{0}|},
\label{bound}
\end{equation}
implying that for finite $\overline{\Psi}_{0}$ is always bounded 
from above so that, for the smallest value of $\Psi_{0}$ allowed in \cite{SH}, 
i.e. $\overline{\Psi}_{0} \sim -1/4$,  $a \leq \frac{4}{3}$. Therefore, 
the expansion only holds locally in time  around $ t\sim t_{0}$. For 
$t\gg t_{0}$ higher gradients (and higher powers of the scale factor) 
must appear. This point has also been correctly emphasized in \cite{selhir} and 
we checked it both by going to higher order in perturbation theory and by extending the present analysis to the other degrees of freedom present in $\gamma_{ij}$ together with $\Psi$. 
 
In the limit $a\to \infty$ the terms neglected in the solution
given in Eqs. (\ref{ans}) and (\ref{sol1a}) become important and the 
momentum constraint is not satisfied anymore. The limit $ a \to \infty$ can be only taken, according to Eq. (\ref{bound}), if $ \overline{\Psi}_{0}\to 0$. But, in this case $q=1/2$. 

To complete the previous series of statements the correct evaluation 
of the deceleration parameter will now be discussed.
Since ${\rm Tr}K^2$ (appearing in the denominator of Eq. (\ref{defq})) must be positive semi-definite, its expression, within the approximations 
made in the derivation of Eq. (\ref{par}), is 
\begin{equation}
{\rm Tr}K^2 = H^2( 1 - 2 a \overline{\Psi}_{0} ),  
\label{correct}
\end{equation}
with $\overline{\Psi}_{0} a < 1/2$; the latter condition is 
less restrictive than the one of Eq. (\ref{limit}). 
Therefore, using the observations reported above (and in particular 
of Eq. (\ref{correct})), the deceleration parameter can be determined 
from Eq. (\ref{defq}) as
\begin{equation}
q(\vec{x},t) = -1 + \frac{3/2 - a \overline{\Psi}_{0}/2}{ ( 1 - 2 a \overline{\Psi}_{0})} = \frac{3 a \overline{\Psi}_{0}  + 1}{2 ( 1 - 2 a \overline{\Psi}_{0} )}.
\label{correctq}
\end{equation}
Taking now into account Eq. (\ref{limit}) we can easily see, from Eq. (\ref{correctq}), that $q(\vec{x},t)$ as a function of  $ a\overline{\Psi}_{0}$ 
is always positive semi-definite. It is amusing that $q(\vec{x},t) \to 0$ for the 
the largest (negative) value of $\overline{\Psi}_{0} a$, i.e. $-1/3$. 
This strongly suggests that the large time limit of $q(\vec{x},t)$ for a maximally underdense Universe (still compatible, though, with perturbation theory) is $0$ 
 as argued in a specific (but exact) example by the authors of Ref. \cite{selhir}.
 
Since the energy density is, to leading order in $|e^{2 f} \nabla^2 f|$
\begin{equation}
\rho =\overline{\rho} + \tilde{\rho} = \frac{3\, H^2}{8\pi G} [ 1 + 3 a \overline{\Psi}_{0}],
\label{endensity}
\end{equation}
when $q(\vec{x}, t) \to 0$, $\rho$ and $\tilde{\rho}$ exactly cancel (as it follows, in the case $p=0$,  from Eq. (\ref{COND3})). 
Thus, for $| a\overline{\Psi}_{0}| \leq 1/3$, $\rho \geq 0$ and all the 
energy conditions are correctly enforced.
 Equation (\ref{endensity}) may also be derived from Eq. 
(\ref{COV0}). To leading order in $|e^{2f} \nabla^2 f|$ (neglecting 
terms $(\nabla f)^2$ and implementing the momentum constraint), Eq. 
(\ref{COV0}) implies:
\begin{equation}
\dot{\overline{\rho}} + 3 H \overline{\rho} =0,\,\,\,\,\,\,\,\,\,\,\,\,
\dot{\tilde{\rho}} + 3 H \tilde{\rho} - 3 \dot{\Psi} \overline{\rho} =0.
\label{COV2}
\end{equation}
Equation (\ref{endensity}) is solution of Eqs. (\ref{COV2}) provided 
terms going as $\dot{\Psi}^2$ are neglected. If this is not the case covariant 
conservation equations and Einstein equations do not lead any longer to 
compatible solutions.
 Once more, if the limit $a\to \infty$ 
is taken blindly in Eq. (\ref{endensity}), $\rho$ becomes arbitrarily 
negative when $\overline{\Psi}_{0} <0$ (underdense regions). But this procedure is not consistent since, at most, $| a \Psi_{0}| < 1/3$ and 
$|\Psi_{0}|<1$.

In \cite{GCA} the authors argue, correctly, that a renormalization of the 
local spatial curvature cannot imply that the Universe is accelerating. Our result shows that this is indeed the case: the acceleration 
cannot be obtained unless the inhomogeneous solution 
is extrapolated in a regime where the terms neglected in the perturbative expansion become dominant. As previously remarked, 
also the results of \cite{selhir}  are consistent both with the present 
treatment and with the criticism raised in \cite{GCA}.  The authors of 
Ref. \cite{selhir} select a specific $f(\vec{x})$ that leads to an exactly 
solvable model with asymptotically vanishing acceleration parameter. 
By comparing the exact result with the proposal of Ref. \cite{SH},  the 
authors of Ref. \cite{selhir}
are led to conclude that the apparent acceleration is just a consequence 
of neglecting higher orders in the gradient expansion. This conclusion
fits well with our findings.
 
Criticisms of the proposal \cite{SH} have been also  presented in \cite{Fla,RAS,siegel}. While these papers raise important issues, they are 
more related with observational implications of a class of solutions whose 
debatable correctness was the aim of the calculations presented here. The interesting points raised in \cite{Fla,RAS,siegel} together with generalizations 
of the arguments presented here will be the subject of a forthcoming publication. 

To conclude, the main lessons to be drawn from the present investigation are
the following: {\it (a)}  in the context of the perturbative 
expansion proposed in \cite{SH} the deceleration parameter
of a matter dominated Universe is always positive to a given 
order in perturbation theory; {\it (b)}  the energy density is always positive 
semi-definite to a given order in perturbation theory;
{\it (c)} the claimed acceleration is the result of the extrapolation of a specific solution in a regime where both, 
the perturbative expansion breaks down and the constraints are violated.

\end{document}